\documentstyle[aps,floats,epsf,color,axodraw]{revtex}
\begin{document}
\baselineskip=14pt
\def\be{\begin{equation}}
\def\ee{\end{equation}}
\def\bea{\begin{eqnarray}}
\def\eea{\end{eqnarray}}
\def\E{{\rm e}}
\def\bearst{\begin{eqnarray*}}
\def\eearst{\end{eqnarray*}}
\def\peleven{\parbox{11cm}}
\def\peffec{\peight{\bearst\eearst}\hfill\peleven}
\def\pspace{\peight{\bearst\eearst}\hfill}
\def\ptwelve{\parbox{12cm}}
\def\peight{\parbox{8mm}}

\title
{Explicit Renormalization Group for D=2 random bond Ising model with
long-range correlated disorder}

\author
{M. A. Rajabpour \footnote{e-mail: rajabpour@physics.sharif.edu},
R. Sepehrinia \footnote{e-mail: sepehrinia@physics.sharif.edu}}


\address
{\it {Department of Physics, Sharif University of Technology, Tehran
,Iran 11365-9161\\}}
\maketitle

\begin{abstract}\baselineskip=12pt
We investigate the explicit renormalization group for fermionic
field theoretic representation of two-dimensional random bond Ising
model with long-range correlated disorder. We show that a new fixed
point appears by introducing a long-range correlated disorder. Such
as the one has been observed in previous works for the bosonic
($\varphi^4$) description. We have calculated the correlation length
exponent and the anomalous scaling dimension of fermionic fields at
this fixed point. Our results are in agreement with the extended
Harris criterion derived by Weinrib and Halperin.
\end{abstract}

\section{introduction}

Critical properties of systems with short-range and long-range
correlated randomness have been studied
extensively\cite{BY,KR,C,br,KLS,h,WH,Elka}. One important question
to address is whether the introduction of weak randomness changes
the universality class of transition. That is, in the
renormalization group (RG) language, whether the disorder is
relevant at the critical point of pure system or not. According to
the well known Harris criterion \cite{C,h}, disorder is irrelevant
if $d \nu>2$, where $d$ is the dimensionality and $\nu$ is the
correlation length exponent of pure system. This criterion should be
modified in the presence of long-range correlations in the disorder.
A special type of such a disorder has been considered by Weinrib and
Halperin. They showed that the disorder with power law correlation
$\sim x^{2\rho-d}$ ( for large separations $x$ ) is irrelevant if
$[(d-2\rho)\nu-2]>0$ for $\rho>0$, whereas the usual Harris
criterion recovers for $\rho<0$ \cite{WH}. Therefore the existence
of long-range correlations in the disorder would have significant
effect in the sense that they can change the universality class of
phase transitions.

The most useful method to study disordered systems in RG language is
the Replica method. By employing this method, one can average out
the free energy using a trick based on the equation $\ln Z = \lim_{n
\rightarrow 0}\frac{Z^n-1}{n}$. The idea is then to average $Z^n$.
However, RG analysis can be implemented explicitly without averaging
on disorder\cite{M}. Using this method, one can avoid some of the
mathematical problems e.g. the $n \rightarrow 0$ limit in the
replica method. Moreover, application of the explicit method in some
cases, such as the one studied in this letter, is more
straightforward.

In this paper we consider random-bond Ising model with fermionic
action and long-range correlated disorder. The effect of short-range
correlated disorder has been studied in previous works
\cite{HL,GL,dd,L,DPP,M}. Also the $\varphi^{4}$ version of this
model is investigated through the double expansion near four
dimensions with short-range and long-rang disorder\cite{WH,Elka}. We
found a new long-range fixed point and we calculate both correlation
length exponent and scaling dimension of fermionic field at this
fixed point.

\section{Explicit Renormalization Group}

Two dimensional Ising model near its critical point can be described
by the massive free fermionic action
\begin{equation}
S=\frac{1}{2}\sum_{x}\bar{\psi}(x)(\hat{\partial}+m)\psi(x),
\end{equation}
$\psi=\tiny{\left(\begin{array}{c}\psi_1 \\ \psi_2
\end{array} \right)}$ is a two component Grassmannian field
$(\psi_i^{\dag}=\psi_i)$ and
\begin{equation}
\bar{\partial}=\sigma_3\partial_1+\sigma_1\partial_2, \hspace{0.5cm}
\bar{\psi}=\psi^{T}i\sigma_2,
\end{equation}
where $\sigma_i$ are Pauli matrices. The two point function of the
Grassmannian fields of the model in the momentum space is
\begin{equation}
G_0(p)(2\pi)^{2}\delta(p+q)=<{\psi}(p){\psi}(q)>
=\frac{-i\hat{p}+m}{p^{2}+m^{2}}(2\pi)^{2}\delta(p+q),
\end{equation}
where ${\psi}(p)$ is the Fourier transform of $\psi(x)$
\begin{equation}
{\psi}(p)=\sum_{x}\psi(x)exp(-ip\cdot x)
\end{equation}
and
\begin{equation}
\psi(x)=\int_{-\pi}^{\pi}\int_{-\pi}^{\pi}\frac{d^{2}p}{(2\pi)^{2}}{\psi}(p)exp(ip\cdot
x).
\end{equation}
The randomness can be inserted into the action in the following way
\begin{equation}
S=\frac{1}{2}\sum_{x}\bar{\psi}(x)(\hat{\partial}+m+c(x))\psi(x),
\end{equation}
where c(x) is a random variable with the following correlation
function in the momentum space
\begin{eqnarray}\label{correlation}
&<&c(p)>=0 \nonumber\\&<&c(p)c(q)>=(D_0+D_\rho|p|^
{-2\rho})\delta(p+q).
\end{eqnarray}

Here $D_0$ and $D_\rho$ are short-range and long-range disorder
strengths respectively. Positivity of two point function imposes
some restrictions on $D_{0}$ and $D_{\rho}$ for example $D_{\rho}$
should be positive for $\rho>0$ and $D_{0}$ should be positive for
$\rho<0$. Here we consider the case with  $0<\rho<1$.\\

We can write the action in the momentum space as
\begin{eqnarray}
S=\frac{1}{2}\int\int_{-\pi}^{\pi}\frac{d^{2}p}{(2\pi)^{2}}\bar{\psi}(-p)
(i\hat{p}+m){\psi}(p)
+\frac{1}{2}\int\int\int\int_{-\pi}^{\pi}\frac{d^{2}p_{1}}{(2\pi)^{2}}
\frac{d^{2}p_2}{(2\pi)^2}\bar{\psi}(p_1)c(-p_{1}-p_{2})\psi(p_{2}),
\end{eqnarray}
where
\begin{eqnarray*}
c(p)=\sum_{x}c(x)exp(-ip.x).
\end{eqnarray*}

It is convenient to introduce a diagrammatic representation. We have
the following vertex
\begin{center}
  \unitlength 1mm
\begin{picture} (80,15)(0,0)
 \linethickness {0.2mm}
 \Photon(10,10)(35,10){1}{6}
 \Line(35,10)(60,10)
 \ZigZag(35,10)(35,25){1.3}{7}
\put(5,0){\small$p_1$}\put(17,0){\small$p_2$}
\put(8,10){\small{$p_1+p_2$}}
\put(23,3){\small$:=\bar{\psi}(p_1)c(-p_1-p_2)\psi(p_2)$.}
\end{picture}
 \end{center}
Wavy line stands for {\small$\bar{\psi}$}, solid line for
{\small$\psi$} and zigzag represents the $c$  insertion. Momentum
conservation should be regarded at each vertex. So we have the
following graphs for vertex renormalization

\begin{center}
  \unitlength 1mm
\begin{picture} (120,15)(0,0)
 \linethickness {0.2mm}
 \Photon(10,10)(30,10){1}{6}
 \DashLine(30,10)(50,10){3}
 \ZigZag(30,10)(30,25){1.3}{7}
 \ZigZag(50,10)(50,25){1.3}{7}
 \Line(50,10)(70,10)
 \put(30,0){\Photon(10,10)(30,10){1}{6}
 \DashLine(30,10)(50,10){3}
 \ZigZag(30,10)(30,25){1.3}{7}
 \ZigZag(50,10)(50,25){1.3}{7}
 \DashLine(50,10)(70,10){3}\ZigZag(70,10)(70,25){1.3}{7}
 \Line(70,10)(90,10)}
\put(70,0){\Photon(10,10)(30,10){1}{6}
 \DashLine(30,10)(50,10){3}
 \ZigZag(30,10)(30,25){1.3}{7}
 \ZigZag(50,10)(50,25){1.3}{7}
 \DashLine(50,10)(70,10){3}\ZigZag(70,10)(70,25){1.3}{7}
 \DashLine(70,10)(90,10){3}\ZigZag(90,10)(90,25){1.3}{7}\Line(90,10)(110,10)}
\put(0,3){$2$}\put(27,3){$+6$} \put(66,3){$+24$}
\put(116,3){$+\cdots$}
\end{picture}
 \end{center}
where the dashed line represents {\small$\bar{\psi}\psi$}
propagator. Symmetry factors will be canceled with $n!$ of
perturbation expansion in each order. The key equation will be
\cite{M}
\begin{eqnarray}\label{key}
c'(r_1,r_2)=c(r_1+r_2)-\int_q
c(r_1-q)G_0(q)c(q+r_2)+\int_{q_1}\int_{q_2}
c(r_1-q_1)G_0(q_1)c(q_1-q_2)G_0(q_2)c(q_2+r_2)\nonumber \\ -
\int_{q_1}\int_{q_2}\int_{q_3}c(r_1-q_1)G_0(q_1)c(q_1-q_2)G_0(q_2)c(q_2-q_3)G_0(q_3)c(q_3+r_2).
\end{eqnarray}

The random function in the original action has zero mean and we
should keep the mean value fixed after RG transformation. So the
mean value of new function should be extracted and in fact it will
renormalize the mass and the kinetic term. Then we need to introduce
a field renormalization constant $Z$ to keep the coefficient of
kinetic term equal to $\frac{1}{2}$, just as in the original action.
After rescaling momentums ($r\rightarrow\lambda r$) and fields
($\psi\rightarrow\frac{\lambda^{3/2}}{\sqrt{Z}} \psi$), the
renormalized random function and mass will be
\begin{eqnarray}
c_r(p_1,p_2)=\frac{1}{\lambda
Z}[c'(p_1/\lambda,p_2/\lambda)-<c'(p_1/\lambda,p_2/\lambda)>].
\end{eqnarray}
\begin{eqnarray}
m_r=\frac{\lambda}{Z}(m + \langle c'\rangle_{r=0})
\end{eqnarray}

A simple expansion of the second term of (\ref{key}) in powers of
$r$, with $m=0$, leads to the following expression for field
renormalization constant to first order in disorder strength

\begin{eqnarray}
Z = 1 + \frac{D_\rho}{4\pi^{1+2\rho}}(\lambda^{2\rho}-1) + O(D^2).
\end{eqnarray}
The short-range correlated disorder does not contribute up to this
order.

So the renormalized mass is
\begin{eqnarray}
m_r=\lambda m\left\{1-\frac{1}{2\pi}(\ln
\lambda)D_0-\frac{1+\rho}{4\rho\pi^{1+2\rho}}(\lambda^{2\rho}-1)D_\rho+((\frac{\ln
 \lambda}{2\pi})^{2}-I_{1})D_{0}^{2}
-(I_{2}+I_{3}+I_{5})D_{0}D_{\rho}-(I_{4}+I_{6})D_{\rho}^{2}+\cdots
\right\},
\end{eqnarray}
where the $I_{j}$'s are
\begin{eqnarray}
I_{1}&=&\int_{q_1}\int_{q_2}\frac{-q_{2}.(q_{2}-q_{1})+m^{2}}{(q_{1}^{2}+m^{2})(q_{2}^{2}+m^{2})((q_{2}-q_{1})^{2}+m^{2})}
\hspace{1cm}|q_{1}-q_{2}|\geq \frac{\pi}{\lambda},\nonumber\\
I_{2}&=&\int_{q_1}\int_{q_2}\frac{(-q_{1}.(q_{2}-q_{1})+m^{2})|q_{1}|^{-2\rho}}{(q_{1}^{2}+m^{2})(q_{2}^{2}+m^{2})((q_{2}-q_{1})^{2}+m^{2})} \hspace{1cm}|q_{1}-q_{2}|\geq \frac{\pi}{\lambda},\nonumber\\
I_{3}&=&\int_{q_1}\int_{q_2}\frac{(-q_{1}.q_{2}+m^{2})|q_{2}-q_{1}|^{-2\rho}}{(q_{1}^{2}+m^{2})(q_{2}^{2}+m^{2})((q_{2}-q_{1})^{2}+m^{2})} \hspace{1cm}|q_{1}-q_{2}|\geq \frac{\pi}{\lambda},\nonumber\\
I_{4}&=&\int_{q_1}\int_{q_2}\frac{(-q_{1}.(q_{2}+q_{1})+m^{2})|q_{1}|^{-2\rho}|q_{2}-q_{1}|^{-2\rho}}{(q_{1}^{2}+m^{2})(q_{2}^{2}+m^{2})((q_{2}-q_{1})^{2}+m^{2})}\hspace{7mm}|q_{1}-q_{2}|\geq \frac{\pi}{\lambda},\nonumber\\
I_{5}&=&\int_{q_1}\int_{q_2}\frac{(-q_{1}.(q_{1}+2q_{2})+m^{2})|q_{2}-q_{1}|^{-2\rho}}{(q_{1}^{2}+m^{2})^2(q_{2}^{2}+m^{2})},\nonumber\\
I_{6}&=&\int_{q_1}\int_{q_2}\frac{(-q_{1}.(q_{1}+2q_{2})+m^{2})|q_{2}-q_{1}|^{-2\rho}|q_{1}|^{-2\rho}}{(q_{1}^{2}+m^{2})^2(q_{2}^{2}+m^{2})}.\nonumber
\end{eqnarray}
we did not include the contribution of $Z$ in the second order.

The renormalized values of disorder strengths can also be obtained
by calculating the correlations of renormalized random function
$c_r(p_1,p_2)$. Up to second order of bare strengthes, the
renormalized strengthes are
\begin{eqnarray}
D_{0r}&=&D_{0}(1-\frac{\ln \lambda}{2\pi}D_{0}-\frac{1+\rho}{2\rho
\pi^{1+2\rho}}(\lambda^{2\rho}-1)D_{\rho}+\cdots),\\
D_{\rho r}&=&\lambda^{2\rho}D_{\rho}(1-\frac{\ln
\lambda}{2\pi}D_{0}-\frac{1+\rho}{2\rho
\pi^{1+2\rho}}(\lambda^{2\rho}-1)D_{\rho}+\cdots).
\end{eqnarray}
By differentiating these equations with respect to $\ln \lambda$,
and then replacing the bare parameters in terms of renormalized
ones, one can obtain the following Wilson's functions
\begin{eqnarray}
\frac{dm_r}{d\ln \lambda}&=&m_{r}(1-\frac{1}{2\pi}D_{0r}-\frac{1+\rho}{2\pi^{1+2\rho}}D_{\rho r}+\cdot\cdot\cdot),\\
\frac{dD_{0r}}{d\ln
\lambda}&=&-D_{0r}(\frac{1}{\pi}D_{0r}+\frac{1+\rho}{\pi^{1+2\rho}}D_{\rho
r}+\cdot\cdot\cdot),\\ \frac{dD_{\rho r}}{d\ln \lambda}&=& D_{\rho
r}(2\rho-\frac{1}{\pi}D_{0r}-\frac{1+\rho}{\pi^{1+2\rho}}D_{\rho
r}+\cdot\cdot\cdot).
\end{eqnarray}

It is clear from the equations that at the Gaussian fixed point
($D_{0}^{*}=D^{*}_{\rho}=0$), $D_{0}$ is marginally irrelevant and
$D_{\rho}$ is relevant. So, by introduction of small amount of
long-range correlated disorder it becomes unstable in $D_{\rho}$
direction. Apart from the trivial Gaussian fixed point, we see that
there is a nontrivial fixed point at $D_{0}^{*}=0$ and
${D^{*}_{\rho}}=\frac{2\rho}{1+\rho}\pi^{1+2\rho}$. The new fixed
point is attractive in all directions in the $D_{0}$ and $D_{\rho}$
plane. The RG flows starting in the vicinity of Gaussian fixed
point, end up at the nontrivial fixed point. One of the most
important critical exponents is the correlation length exponent,
which turns out to be different at two fixed points, $\nu=1$ at the
Gaussian fixed point and $\nu=\frac{1}{1-\rho}$ at the nontrivial
fixed point. The result is in agreement with \cite{WH}. A good
numerical confirmation of this relation for the special choice of
$\rho=\frac{1}{2}$ can be found in \cite{Igloi}.

Also the field renormalization would acquire an anomalous dimension
which is defined through the asymptotic behavior of vertex function
in the long wavelength limit as

\begin{equation}
\Gamma^{(2)}(p)\sim p^{1-\eta}
\end{equation}
so at the nontrivial fixed point we have
\begin{equation}
\eta=\frac{dZ}{d\ln\lambda}=\frac{\rho}{2\pi^{1+2\rho}}D^*_{\rho
r}=\frac{\rho^2}{1+\rho}
\end{equation}

Now we want to show that, some less singular terms in the
correlation functions of $c_{r}$, which were omitted in the equation
(\ref{correlation}), are irrelevant. We start with a general form of
the correlation function of disorder
\begin{eqnarray}
&<&c(p)>=0,
\nonumber\\&<&c(p)c(q)>=(D_0+\sum_{i=1}^{n}D_{\rho_{i}}|p|^
{-2\rho_{i}})\delta(p+q),
\end{eqnarray}
and then find the Wilson's functions for the $D_0$ and
$D_{\rho_{i}}$
\begin{eqnarray}
\frac{dD_{0r}}{d\ln\lambda}&=&-D_{0r}(\frac{1}{\pi}D_{0r}+\sum_{i=1}^{n}\frac{1+\rho_i}{\pi^{1+2\rho_{i}}}D_{\rho_{i}
r}+\cdot\cdot\cdot),\\
\frac{dD_{\rho_{i} r}}{d\ln \lambda}&=& D_{\rho_{i}
r}(2\rho_i-\frac{1}{\pi}D_{0r}-\sum_{j=1}^{n}\frac{1+\rho_j}{\pi^{1+2\rho_{j}}}D_{\rho_{j}
r}+\cdot\cdot\cdot).
\end{eqnarray}

Here we have $n$ nontrivial fixed points at $D_{0}^{*}=0$ and all
$D^{*}_{\rho_{i}}=0$ except one of them, say ${D^{*}_{\rho_{l} }}$
which is $\frac{2\rho_{l}}{1+\rho_l}\pi^{1+2\rho_{l}}$. At these
fixed points the RG eigenvalues are $-2\rho_{l}$ and
$2(\rho_{i}-\rho_{l})$ which means that there is just one stable
fixed point (with $\rho_{l}>\rho_{i}$ for all $i$). Other fixed
points are unstable at least in one direction.

Finally  we want to compare the above results with the replica
results. In the case of short-range disorder, as pointed out by
Murthy\cite{M}, the results of the replica method and the explicit
method are in complete agreement. The replica action for the model
considered here can be easily obtained,
\begin{eqnarray}
S=\frac{1}{2}\sum_{x,\alpha}\bar{\psi}(x)(\hat{\partial}+m)\psi(x)-\frac{D_{0}}{8}
\sum_{x,\alpha,\beta}\bar{\psi}_{\alpha}(x)\psi_{\alpha}(x)\bar{\psi}_{\beta}(x)\psi_{\beta}(x)-\nonumber\\
\frac{D_{\rho}}{8}\sum_{x,y,\alpha,\beta}\bar{\psi}_{\alpha}(x)\psi_{\alpha}(x)\frac{1}{|x-y|^{2-2\rho}}\bar{\psi}_{\beta}(y)\psi_{\beta}(y).
\end{eqnarray}

The third term in this action is nonlocal, and the computation of
the $\beta$ functions are not easily tractable. This shows the
advantage of explicit calculations with which one can avoid such a
nonlocal action. The other advantage of this method is that there is
no restriction on the distribution function of disorder while it
should have a Gaussian distribution in the replica method.

\section{Acknowledgment}
We thank Professor G. Murthy for helpful discussions and Professor
J. Cardy for useful comments. R. S. would like to thank E. Khatami
and A. A. Saberi for critical reading of manuscript.

\end{document}